\documentclass[journal = nalefd, manuscript = letter]{achemso}

\usepackage[numbers]{natbib}
\usepackage{amsmath,amssymb}
\usepackage{graphicx}
\usepackage{dcolumn}
\usepackage{bm}
\usepackage{xcolor}

\author{Petr Stepanov}
\affiliation{Univ. Grenoble Alpes, F-38000 Grenoble, France}
\alsoaffiliation{CEA, INAC-PHELIQS, ``Nanophysique et semiconducteurs" group, F-38000 Grenoble, France}

\author{Marta Elzo Aizarna}
\affiliation{Univ. Grenoble Alpes, F-38000 Grenoble, France}
\alsoaffiliation{CEA, INAC-PHELIQS, F-38000 Grenoble, France}
\alsoaffiliation{ESRF-The European Synchrotron, 38043 Grenoble Cedex 9, France}

\author{Jo{\"e}l Bleuse}
\affiliation{Univ. Grenoble Alpes, F-38000 Grenoble, France}
\alsoaffiliation{CEA, INAC-PHELIQS, ``Nanophysique et semiconducteurs" group, F-38000 Grenoble, France}

\author{Nitin S. Malik}
\affiliation{Univ. Grenoble Alpes, F-38000 Grenoble, France}
\alsoaffiliation{CEA, INAC-PHELIQS, ``Nanophysique et semiconducteurs" group, F-38000 Grenoble, France}

\author{Yoann Cur\'{e}}
\affiliation{Univ. Grenoble Alpes, F-38000 Grenoble, France}
\alsoaffiliation{CEA, INAC-PHELIQS, ``Nanophysique et semiconducteurs" group, F-38000 Grenoble, France}

\author{Eric Gautier}
\affiliation{Univ. Grenoble Alpes, F-38000 Grenoble, France}
\alsoaffiliation{CEA-CNRS, INAC-SPINTEC, F-38000 Grenoble, France}

\author{Vincent Favre-Nicolin}
\affiliation{Univ. Grenoble Alpes, F-38000 Grenoble, France}
\alsoaffiliation{CEA, INAC-PHELIQS, F-38000 Grenoble, France}
\alsoaffiliation{Institut Universitaire de France, Paris, France}

\author{Jean-Michel G{\'e}rard}
\email{jean-michel.gerard@cea.fr}
\affiliation{Univ. Grenoble Alpes, F-38000 Grenoble, France}
\alsoaffiliation{CEA, INAC-PHELIQS, ``Nanophysique et semiconducteurs" group, F-38000 Grenoble, France}

\author{Julien Claudon}
\affiliation{Univ. Grenoble Alpes, F-38000 Grenoble, France}
\alsoaffiliation{CEA, INAC-PHELIQS, ``Nanophysique et semiconducteurs" group, F-38000 Grenoble, France}

\title{Large and uniform optical emission shifts in quantum dots externally strained along their growth axis}

\date{\today}

\begin{document}

\hyphenation{micro-photo-lumi-nescence}

\begin{abstract}

We introduce a method which enables to directly compare the impact of elastic strain on the optical properties of distinct quantum dots (QDs). Specifically, the QDs are integrated in a cross-section of a semiconductor core wire which is surrounded by an amorphous straining shell. Detailed numerical simulations show that, thanks to the mechanical isotropy of the shell, the strain field in a core section is homogeneous. Furthermore, we use the core material as an {\it in situ} strain gauge, yielding reliable values for the emitter energy tuning slope. This calibration technique is applied to self-assembled InAs QDs submitted to incremental tensile strain along their growth axis. In contrast to recent studies conducted on similar QDs stressed perpendicularly to their growth axis, optical spectroscopy reveals $5-10$ times larger tuning slopes, with a moderate dispersion. These results highlight the importance of the stress direction to optimise QD response to applied strain, with implications both in static and dynamic regimes. As such, they are in particular relevant for the development of wavelength-tunable single photon sources or hybrid QD opto-mechanical systems.


\end{abstract}


\maketitle

Applying an external strain field on a semiconductor quantum dot (QD) is a powerful method to tailor its optical properties without compromising its brightness. In the last years, elastic tuning has been exploited to induce shifts in the QD emission energy~\cite{Ding_PRL_10,Jons_PRL_11,Kuklewicz_NanoLett_12,Bavinck_NanoLett_12}, enabling the realization of wavelength-tunable single-photon sources~\cite{Wu_APL_13,Kremer_PRB_14}. Such a `tuning knob' thus allows bringing distinct QDs into resonance~\cite{Flagg_PRL_10}, a basic requirement to realize a quantum photonic circuit. Furthermore, full control over the in-plane strain tensor can compensate the natural asymmetry of as-grown QDs~\cite{Trotta_PRL_15}, with important application to the deterministic emission of polarization-entangled photon pairs~\cite{Chen_NatComm_16,Trotta_NatComm_16}. Strikingly, the nature of the excitonic ground state hosted by a QD can be controlled by an external bi-axial tensile strain~\cite{Huo_NatPhys_14}. In the dynamical regime, surface acoustic waves open a route to manipulate the QD emission wavelength on short time scales~\cite{Gell_APL_08,Metcalfe_PRL_10,Weiss_NanoLett_14,Weiss_JPD_14,Schulein_NatNano_15}. Finally, elastic coupling lies at the heart of recently demonstrated QD opto-mechanical hybrid systems~\cite{Yeo_NatNano_14,Montinaro_NanoLett_14}, which enable new approaches to precision sensing or to the exploration of the quantum-classical boundary~\cite{Treutlein_14}. 

Beside single QD demonstrations, exploring the distribution of the strain-induced optical shifts in a QD ensemble combines fundamental and practical interests. Indeed, such non-destructive studies offer a valuable insight in the dot structural properties~\cite{Jons_PRL_11,Kuklewicz_NanoLett_12}. Moreover, a calibration of the strain response constitute an important input for the design of QD devices exploiting strain tuning, while a moderate dispersion of QD properties is highly desirable to allow a reasonable fabrication yield. To enable a direct dot-to-dot comparison, distinct emitters should ideally experience an identical and well characterized strain field. Achieving the required strain uniformity is generally challenging. 

In this context, it was shown recently that the emission energy of a single QD embedded in a nanowire can be largely tuned by a dielectric straining shell~\cite{Bavinck_NanoLett_12}. We furthermore demonstrate here that an amorphous shell induces a uniform strain field in the crystalline core, even if the latter features a pronounced mechanical anisotropy. The semiconductor core experiences a longitudinal strain, whose magnitude is determined by measuring the core absorption spectrum. As a consequence, one then directly obtains the tuning slope of several QDs embedded in a core cross-section. This calibration technique is applied to self-assembled InAs QDs embedded in a GaAs-silica core-shell structure. At low temperature, we track the evolution of individual QD photo-luminescence lines submitted to incremental elongation along their growth axis. In contrast to recent studies on similar QD subject to uniaxial stress perpendicular to the growth axis~\cite{Jons_PRL_11,Kuklewicz_NanoLett_12}, our measurements reveal five times larger tuning slopes, with a moderate dispersion. These results highlight the importance of the stress direction to optimise QD response to applied strain, with implications both in static and dynamic regimes.

 
\begin{figure}
\centering
\includegraphics[width=0.95\textwidth]{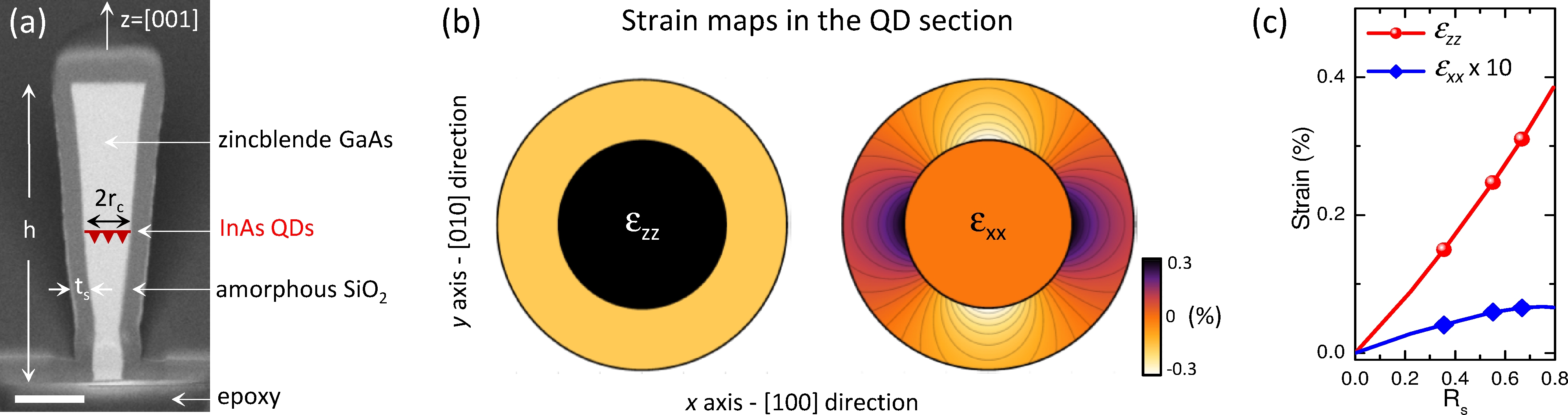}
\caption{{\bf Elastic strain in a hybrid core-shell nanostructure.} (a) Tilted scanning electron microscope view of a fabricated structure cut vertically along a diameter with a focused ion beam system. The longitudinal $z$ axis is aligned along the $[001]$ core crystal direction. Quantum dots (QDs) are pictured as triangles. The scale bar represents $400\: $nm. (b) Calculated maps of $\epsilon_{zz}$ and $\epsilon_{xx}$ in the QD cross-section ($t_s = 110\: $nm). The $\epsilon_{yy}$ map is obtained by rotating the $\epsilon_{xx}$ map by $90^{\circ}$. Inside the core, strain components are uniform. (c) $\epsilon_{zz}$ and $\epsilon_{xx}$ experienced by the QDs, as a function of the core-shell relative area $R_s = 1 - r_c^2/(r_c+t_s)^2$. All calculations are conducted for $r_c = 150\: $nm and for a structure cooled down from $T_d^\text{eff} = 1000\: $K to $T=4\: $K.}
\label{fig:fig1}
\end{figure}

The hybrid core-shell nanostructure investigated in this work is shown in Fig.~\ref{fig:fig1}(a). The core is made of zincblende GaAs; its axis coincides with the $z = [001]$ crystalline direction. It features a height $h = 1.73\: \mu$m and contains a single sheet of self-assembled InAs QDs, located at equal distance from the wire terminations. This slightly conical structure (sidewall angle $\alpha = 6^{\circ}$), was defined by etching a planar structure grown by molecular beam epitaxy (see Methods). After process, this core wire is capped with a conformal amorphous SiO$_2$ layer deposited at $T_d = 550\: $K in a PECVD chamber. Since SiO$_2$ features a much smaller linear thermal expansion coefficient than GaAs, a tensile strain is applied on the core when the structure is cooled down below $T_d$. As shown later, the silica layer also features a significant built-in strain, common for such dielectric layers~\cite{Tarraf_JMM_04}. The total strain applied to the core can empirically reproduced with an effective SiO$_2$ deposition temperature $T_d^\text{eff} = T_d + 450\: $K. 

It is well known that in a cylindrical core-shell nanowire composed of mechanically isotropic materials, the strain field inside the core is uniform far enough from the wire terminations. In contrast, if both core and shell are made of zincblende materials aligned along the $[001]$ direction, mechanical anisotropy induces a pronounced spatial modulation of the strain field inside the core~\cite{Ferrand_TEPJAP_14}. We explore here a hybrid situation: the anisotropic core is strained by an isotropic shell. As detailed in Methods, we compute the strain tensor $\bar{\bar{\epsilon}}$ in the $(x,y,z) = ([100],[010],[001])$ basis, for a structure cooled from $T_d^\text{eff} = 1000\: K$ down to liquid helium temperature ($4\: $K). Calculations are conducted for the fabricated geometry, and we focus on strain distribution in the QD plane. At this location, the core radius $r_c$ is small enough so that finite-length effects can be neglected ($2r_c < h/2$). Inside the core, off-diagonal components of $\bar{\bar{\epsilon}}$ can be neglected and the strain tensor features an in-plane symmetry ($\epsilon_{xx} = \epsilon_{yy}$). In addition, the maps shown in Fig.~\ref{fig:fig1}(b) reveal that all strain components are uniform, with the hierarchy $|\epsilon_{zz}| \gg |\epsilon_{xx}|, |\epsilon_{yy}|$. Therefore, QDs randomly located in the core section will experience an identical elongation along their growth axis. As shown in Fig.~\ref{fig:fig1}(c), $\epsilon_{zz}$ is roughly proportional to the core to shell relative area $R_s = 1 - r_c^2/(r_c+t_s)^2$, where $t_s$ is the shell thickness.


\begin{figure}
\centering
\includegraphics[width=0.45\textwidth]{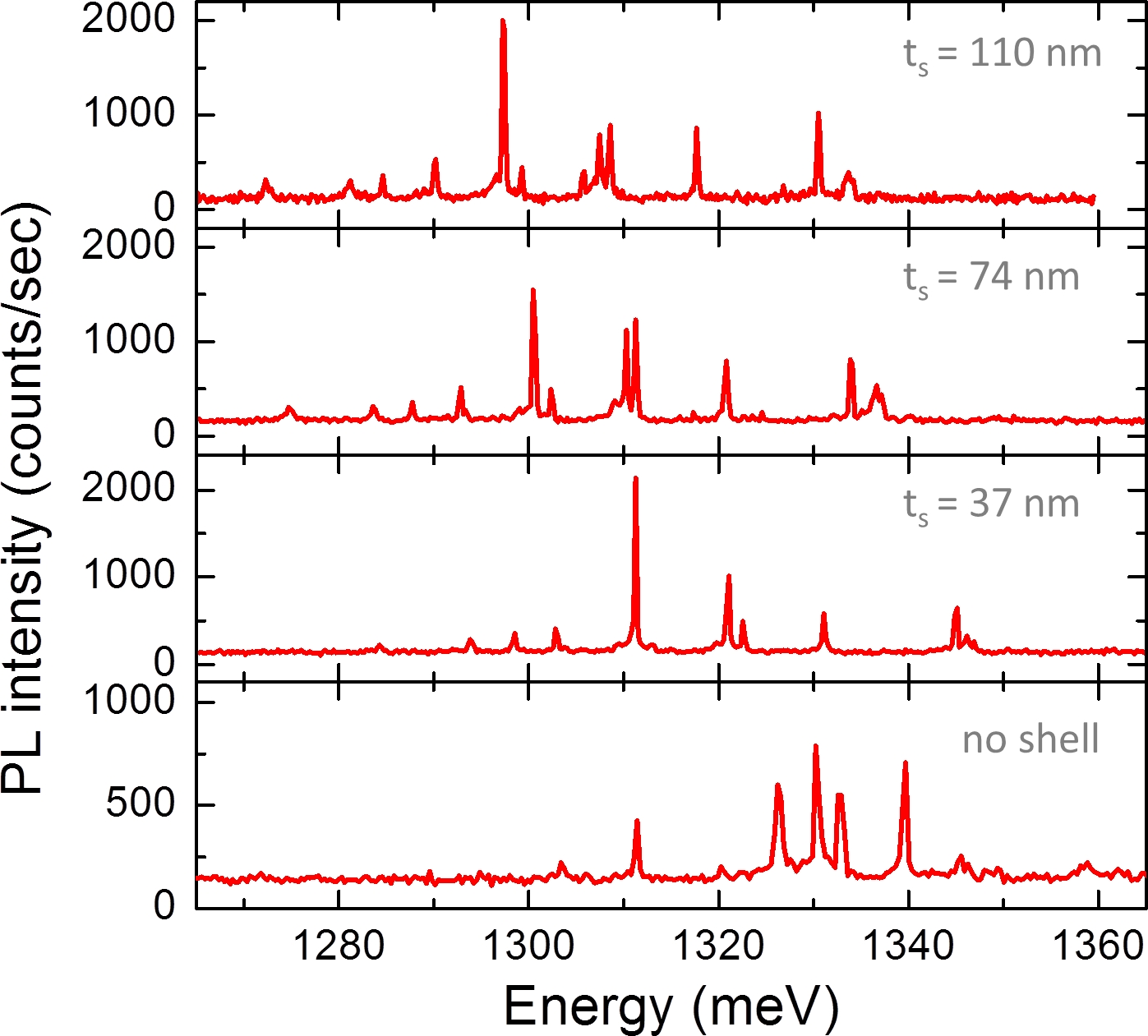}
\caption{{\bf Individual QDs subject to incremental external strain.} Micro-photoluminescence spectra acquired on the same nanowire ($r_c = 150\: $nm) for increasing shell thicknesses $t_s$. Measurements are conducted at $4\: $K. }
\label{fig:fig3}
\end{figure}


We now investigate the optical properties of individual QDs that experience incremental external strain, applied by the successive deposition of `thin' silica shells. Optical characterization is performed in a micro-photoluminescence ($\mu$PL) setup at liquid helium temperature ($T = 4\: $K). Figure~\ref{fig:fig3} shows four spectra, acquired on the same wire ($r_c = 150\: $nm): the bottom one corresponds to the uncapped nanowire, and the three others are obtained after successive deposition of three nominally identical SiO$_2$ layers. Each deposition step covers the sidewalls with an additional shell of thickness $37\: $nm. All spectra feature sharp lines, associated with the recombination of excitonic complexes trapped in QDs. The excitation power $P_\text{ex} \sim 1\: \mu$W is typically half of the saturation power of the brightest lines, ensuring that the spectra are dominated by the recombination of the lowest excitonic complexes (neutral exciton, trion and bi-exciton). Even though we did not conduct a detailed identification of the nature of each line, we stress that the energy span (around $80 \: $meV) is sufficient to ensure that several distinct QDs contribute to the signal.

We first consider the impact of the deposition of the first shell on a bare wire. In that case, it is not possible to reliably follow the evolution of individual spectral lines, essentially because of the appearance of new lines in the capped wire. We attribute this to a change in the semiconductor surface, which could modify the pumping efficiency and/or radiative yield of embedded QDs, especially those which are close to the sidewall. Therefore, we concentrate our analysis on the deposition of the second and third shells, which should not induce further modification of the semiconductor interface. In both cases, the spectrum modification can be roughly described as a rigid redshift of the main spectral features. This constitutes a first indication that the dots embedded in the structure exhibit a similar response to the external strain.

\begin{figure}
\centering
\includegraphics[width=0.45\textwidth]{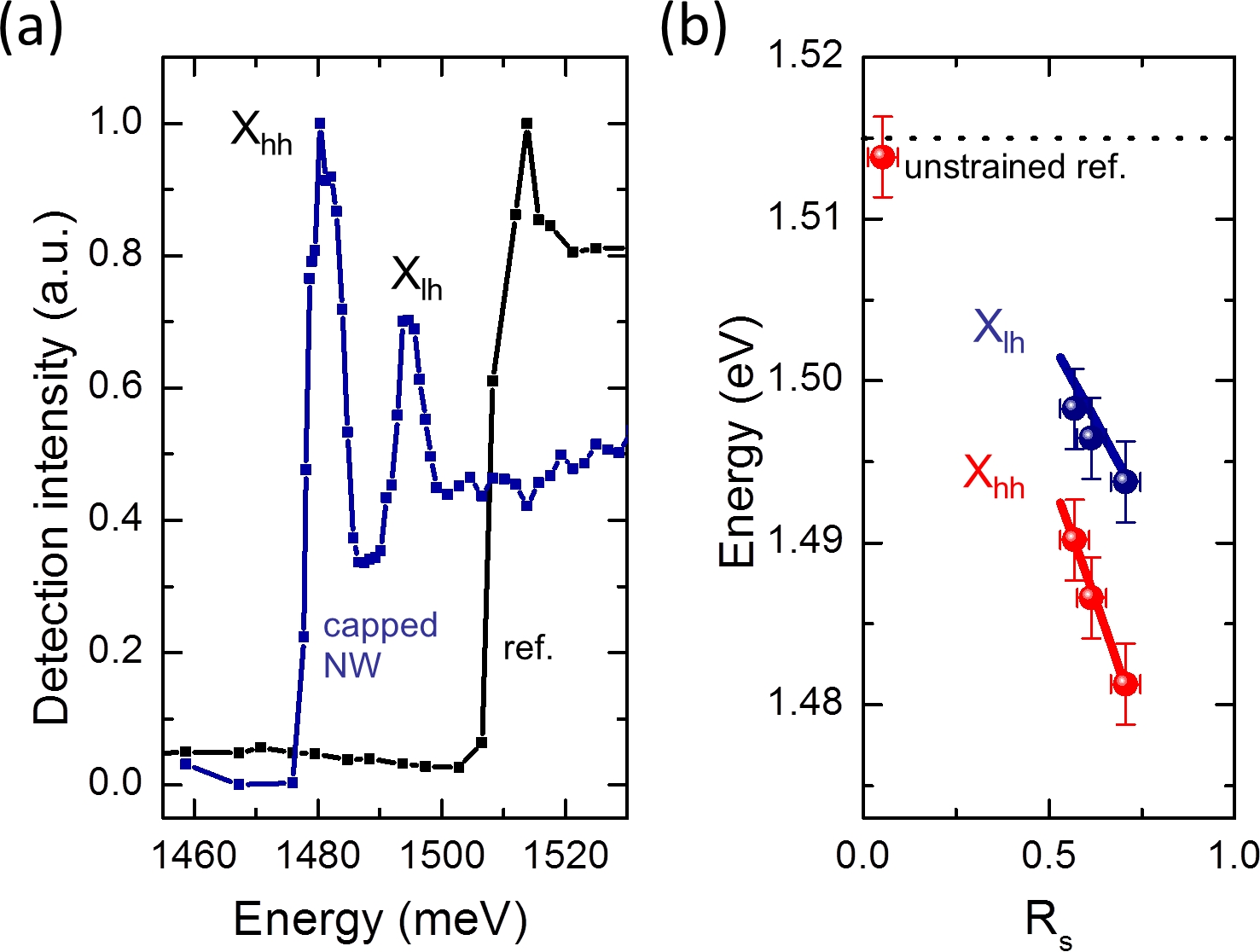}
\caption{{\bf Nanowire core material as an {\it in situ} strain gauge.} (a) Absorption spectrum of GaAs, as measured by a photoluminescence excitation technique, for a nanowire with $r_c = 150\: $nm and $t_s = 110\: $nm. The peaks are attributed to heavy-hole and light-hole free excitons of bulk GaAs, $X_{hh}$ and $X_{lh}$, respectively. The curve `ref.' is acquired on a large, and thus nearly unstrained structure. (b) Solid dots: Peak energies of the heavy- and light-hole free excitons measured on three different wires ($r_c = 130, 150, 180\: $nm and $t_s = 110\: $nm), plotted against $R_s$. `ref.' corresponds to the nearly unstrained reference, close to tabulated energy of the unstrained GaAs free exciton (dashed line). The solid lines are theoretical adjustment with a single free parameter: the effective temperature of the silica deposition, $T_d^\text{eff} = 1000\: $K.}
\label{fig:fig4}
\end{figure}

The main components of the strain field induced by the shell in the core can be determined {\it in situ} by a non-destructive optical measurement. To this end, we use the bulk GaAs core material as a very sensitive strain gauge. Indeed, a relative change in the crystal unit cell volume, proportional to $\epsilon_h = \epsilon_{xx} + \epsilon_{yy} + \epsilon_{zz}$, shifts the bandgap energy. In the core, anisotropic strain components reduce to the tetragonal shear strain $\epsilon_{sh} = 2 \epsilon_{zz} - \epsilon_{xx} - \epsilon_{yy}$. It leaves the conduction band unaffected, but splits the valence states into a heavy hole ({\it hh}) and a light hole ({\it lh}) band~\cite{Bir_1974,Marzin_SS_90}. We access the band structure of strained GaAs by measuring its absorption spectrum with a photoluminescence excitation technique (see Methods). Figure~\ref{fig:fig4}(a) shows two absorption spectra acquired on two different structures covered by a $110\: $nm thick shell. The first one (curve `ref.') features lateral dimensions much larger than $t_s$. As a consequence, the strain induced by the shell is small and the absorption spectrum features a single peak, attributed to the bulk free exciton of GaAs. The peak energy is close to the tabulated value for unstrained GaAs $E_X^0 =1.515\: $eV (Ref.~\cite{Bogardus_PR_68}). The second one is acquired on a nanowire with $r_c = 150\: $nm. The absorption spectrum undergoes a global redshift and features two peaks. The low-(high-) energy transition, centered around $E_{X_{hh}}$ ($E_{X_{lh}}$) is attributed to the  {\it hh} ({\it lh}) free GaAs exciton. Similar measurements were conducted on two other wires: the results appear as solid points in Fig.~\ref{fig:fig4}(b).

Neglecting variation in the exciton binding energy with strain, the mean shift is simply given by $\frac{1}{2}(E_{X_{hh}} + E_{X_{lh}}) - E_X^0 = a \epsilon_h$ and the peak splitting by $E_{X_{lh}} - E_{X_{hh}} = - b \epsilon_{sh}$, where $a = a_c + a_v = -8.33\: $eV and $b = -2.0\: $eV are the deformation potentials of GaAs~\cite{Vurgaftman_JAP_03}. The strain deduced from the peak positions is larger than the one calculated using the thermal dilatation coefficients of GaAs and SiO$_2$, for a structure cooled from $T_d = 550\: $K down to $4\: $K. This evidences an additionnal tensile built-in strain in the shell. We note that such built-in strain is common in dielectric layers deposited by PECVD, and depends on the deposition parameters~\cite{Tarraf_JMM_04}. As illustrated in Fig.~\ref{fig:fig4}(b), the total strain in the core can be empirically reproduced with an effective SiO$_2$ deposition temperature $T_d^\text{eff} = 1000\: $K. In the following, all strain values are calculated using the calibrated $T_d^\text{eff}$ and geometrical parameters obtained from SEM observation. 

\begin{figure}
\centering
\includegraphics[width=0.35\textwidth]{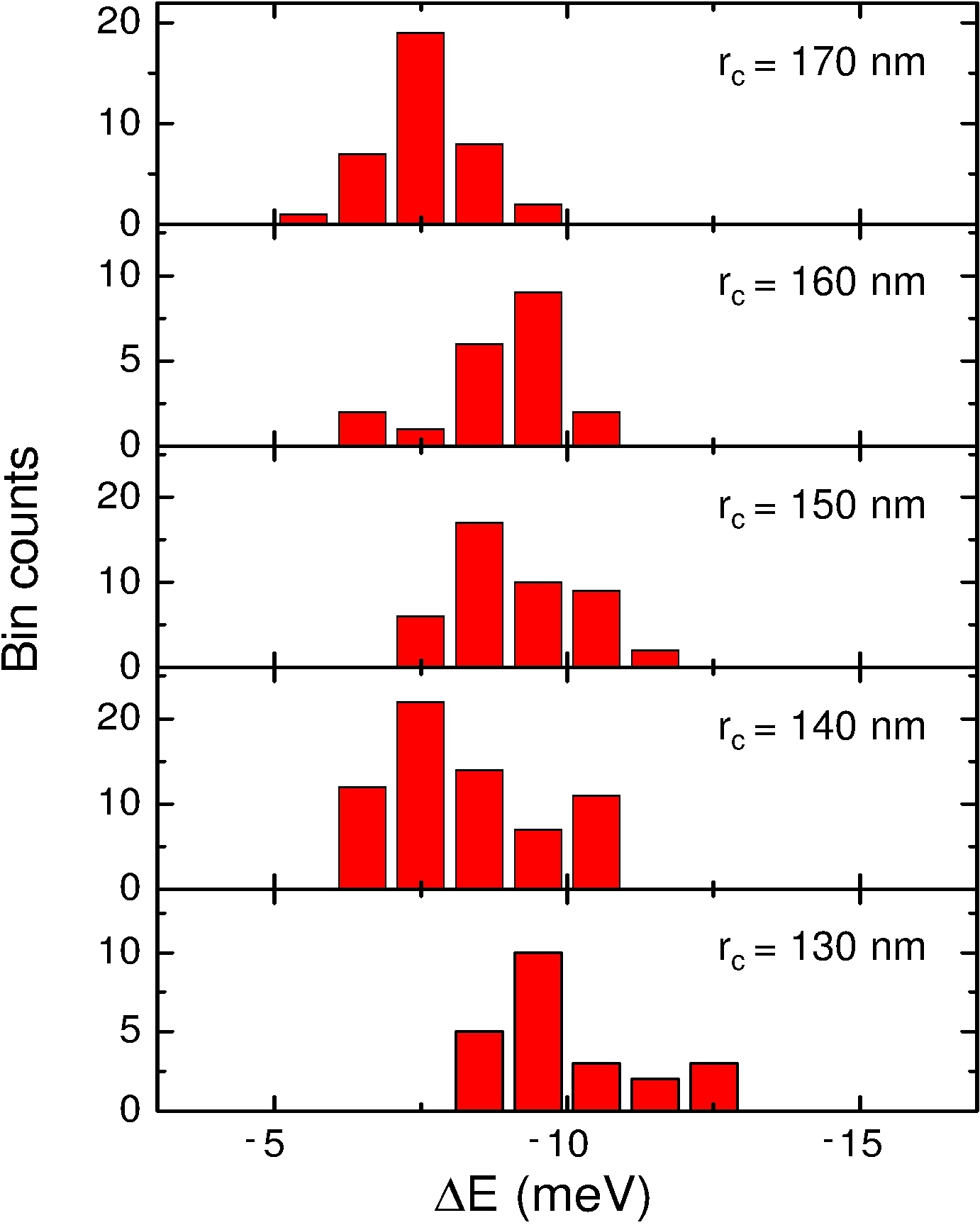}
\caption{{\bf Histograms of the spectral shifts $\Delta E$ of individual QD lines.} Five wire families, characterized by their core radius $r_c$, were investigated. We focus on shifts induced by an increase of the SiO$_2$ shell from $t_s=37\: \text{nm}$ to $73\: \text{nm}$.}
\label{fig:fig5}
\end{figure}

After this calibration, we come back to the analysis of the QD spectral shift $\Delta E$. The histograms in Fig.~\ref{fig:fig5} show the distribution of $\Delta E$ measured in five nanowire families with $r_c$ ranging from $130\: $nm to $170\: $nm, after an increase of the shell thickness from $t_s = 37\: \text{nm}$ to $73 \: $nm. In total, $25$ different nanowires and $190$ spectral lines were investigated. All these lines exhibit a redshift under tensile strain. For each wire family, Table~\ref{tab:tab1} summarizes the mean shift $\left< \Delta E \right>$ and the standard deviation $\sigma_{\Delta E} = [ \left< \Delta E^2 \right> - \left< \Delta E \right>^2]^\frac{1}{2}$. The table also provides $\left< \alpha \right>$ and $\sigma_{\alpha}$, the mean value and standard deviation of the tuning slope $\alpha = \Delta E/\Delta \epsilon_{zz}$ ($\Delta \epsilon_{zz}$ is the increase in tensile strain along $z$). As expected, the mean tuning slopes are similar for all wires, with an average value of $- 91 \: \text{meV}/\%$. Furthermore, the relative dispersion in the tuning slopes is relatively small, with an average $\sigma_{\alpha}/|\left< \alpha \right> | = 13\: \%$.

\begin{table}
\centering
   \begin{tabular}{ c l l l l l l}
		\hline
     $r_c$                                                                          & (nm)               & 130   & 140  & 150   & 160   & 170 \\
		 \hline
		 \hline
		 $\Delta \epsilon_{zz}$																													& ($\%$)						 & 0.10 & 0.098& 0.095 & 0.092 & 0.090  \\
		 $\left< \Delta E \right>$                                                      & (meV)              & -9.8  & -8.2 & -9.0  & -8.8  & -7.6 \\
     $\sigma_{\Delta E}$                                                            & (meV)              & 1.2   & 1.3  & 1.1   & 1.1   & 0.90 \\
		 $\left< \alpha \right> = \frac{\left< \Delta E \right>}{\Delta \epsilon_{zz}}$ & ($\text{meV}/\%$)  & -98   & -84  &  -95  & -96   & -84  \\
		 $\sigma_{\alpha}$                                                              & ($\text{meV}/\%$)  & 12    & 13   &  12   & 12    & 10   \\
     \hline
   \end{tabular}
\caption{Statistical analysis of the histograms shown in Fig.~\ref{fig:fig5}. Strain variation $\Delta \epsilon_{zz}$, mean value ($\left< \Delta E \right>$) and standard deviation ($\sigma_{\Delta E}$) of the energy shift $\Delta E$. Mean value ($\left< \alpha \right>$) and standard deviation ($\sigma_{\alpha}$) of the tuning slope $\alpha = \frac{\Delta E}{\Delta \epsilon_{zz}}$.}
\label{tab:tab1}
\end{table}

There are few reported studies which provide a statistical analysis of the response of different QDs to strain. We present here such a study for the first time for self-assembled InAs QD submitted to a tensile strain along the growth direction [001]. In particular, our results contrast with the ones obtained on self-assembled InGaAs QD submitted to a tensile stress along the [110] directions~\cite{Jons_PRL_11,Kuklewicz_NanoLett_12}. Strikingly, both redshifts and blueshifts were simultaneously observed in both references. In addition to this large dispersion, the mean tuning slopes were typically 5-10 times smaller than the one obtained in this work.

To interpret qualitatively this remarkable discrepancy, we come back to the main effects of external elastic strain on the QD emission energy. To first order, the strain response is dominated by single-particle effects, {\it i.e.} by a modification of single electron and hole energy levels~\cite{Ding_PRL_10,Kuklewicz_NanoLett_12,Wu_APL_13}. The variation in the QD emission energy can then be expressed as $\Delta E = \Delta E_{g,hh} + \Delta E_c$. The first term is the modification of the {\it hh} bandgap of the QD material and the second one is associated with variations in the electronic and hole confinement energies, via a change of the QD height and confinement potentials. External strain adds to the bi-axial compressive strain initially present in the QD. If the external strain field features the same symmetry, $\Delta E_{g,hh} = a \epsilon_h + \tfrac{b}{2} \epsilon_{sh}$, where $a= a_c + a_v$ and $b$ are the (negative) deformation potentials of the QD material. For an elongation parallel to [001] (our work), $\epsilon_h \approx \epsilon_{zz}$ and $\epsilon_{sh} \approx 2 \epsilon_{zz}$. In the case of a pure InAs QD, this yields $\Delta E_{g,hh} = -79 [\text{meV}] \times \epsilon_{zz} [\%]$, which constitutes the leading term for the tuning slope. Confinement effects also contribute to a redshift, with an estimated 5 times smaller amplitude. Though simple, this estimation agrees remarkably well with the observed tuning slopes.

We now turn to the situation investigated in Refs.~\cite{Jons_PRL_11} and  ~\cite{Kuklewicz_NanoLett_12} : the QD experiences a uniaxial tensile {\it stress} applied along [110]. For simplicity, we treat the GaAs membrane as an isotropic material, with a Poisson ratio $\nu = 0.31$. In the basis adapted to the stress configuration $(x'=[110],y'=[1\bar{1}0],z'= z)$, the strain tensor is diagonal, with $\epsilon_{y'y'}=\epsilon_{z'z'} = -\nu \epsilon_{x'x'}$. Coming back to the $(x,y,z)$ crystal basis, $\epsilon_h = (1-2\nu) \epsilon_{x'x'}$ is strongly reduced by the material contraction in transverse directions. Furthermore, $\epsilon_{sh} = -(1 + \nu) \epsilon_{x'x'}$ is now opposite to $\epsilon_{h}$. The external strain in-plane anisotropy introduces an off-diagonal term $\epsilon_{xy} = -\tfrac{1+\nu}{2} \epsilon_{x'x'}$; its contribution to $\Delta E_{g,hh}$ is proportional to $\epsilon_{x'x'}^2$ and can be neglected when the applied strain is smaller than the initial QD bi-axial strain. $\Delta E_{g,hh}$ then takes a similar form as in the previous paragraph, but with a much smaller tuning slope ($\Delta E_{g,hh} = -11 [\text{meV}] \times \epsilon_{x'x'} [\%]$). As a result, the QD response becomes highly sensitive to modifications of the confinement energy, which in turn critically depends on the QD morphology (size, alloy composition). Inconsistent shifts with small amplitude and variable sign are then observed.

For completeness, let us mention that, in general, strain can also induce piezo electric fields~\cite{Weiss_NanoLett_14,Weiss_JPD_14} and thus Stark shifts of the QD emission. In our work, these Stark shifts are typically $10^3$ times smaller than the leading strain tuning term, and thus negligeble. For uniaxial stress applied along [110], they however feature a significant amplitude, 2-3 times smaller than strain-induced shifts .

This analysis highlights the high sensitivity of the QD response to the exact structure of the applied strain field, an important feature for applications both in static and dynamic regimes. Given the hierarchy of the deformation potentials, response to strain is optimized when (i) $\epsilon_h$ is maximized and (ii) the effects of $\epsilon_h$ and $\epsilon_{sh}$ add constructively. One then obtains large spectral shifts, with moderate dot-to-dot variability. These results are general, with implications for all strained-QD devices. We now discuss some important particular cases. In the static regime, digital shell tuning~\cite{Bavinck_NanoLett_12} can be directly applied to efficiently tune QD-photonic wire devices~\cite{Claudon_ChemPhysChem_13,Munsch_PRL_13,Kremer_PRB_14,Versteegh_NatComm_14}. We have shown here that the large, discrete, spectral shifts induced by the shell can be predicted with a simple theory. Therefore, this technique complements continuous methods, which generally feature a limited tuning range. Interestingly enough, the large, controlled bi-axial tensile strain induced by the shell offers a very simple way to stabilize a light-hole ground state in a GaAs QD~\cite{Huo_NatPhys_14}. In the dynamical regime, maximizing the sensitivity to strain is particularly relevant for recently demonstrated hybrid QD opto-mechanical systems, for which periodic lattice deformation couples the QD exciton energy to the displacement of a mechanical oscillator~\cite{Yeo_NatNano_14,Montinaro_NanoLett_14}. Our results show that one should favor geometries in which stress is locally applied along the QD growth axis, such as in flexural or longitudinal vibration modes in $z$-oriented wires~\cite{Yeo_NatNano_14}. In view of the relatively small tuning slopes consistently measured for in-plane uniaxial stress~\cite{Seidl_APL_06,Jons_PRL_11,Kuklewicz_NanoLett_12}, an in-plane cantilever, initially proposed in Ref.~\cite{Wilson-Rae_PRL_04}, seems less favorable to maximize the strain coupling. 

The calibration technique demonstrated in this work is versatile. Indeed, strain uniformity in a zincblende core is maintained for other wire orientations ($z=<110>$ and $<111>$) and also holds for elliptical cross-sections. Furthermore, strain is also uniform in the core for a wurtzite wire defined along the $c$-axis. Therefore, this strategy could be applied to characterize the strain response of QDs or optically active defects in other important material systems. In particular, determining the strain response of individual spins of charge carriers~\cite{Gao_NatPhot_15} or single dopants~\cite{Besombes_Nanophotonics_15} in QDs is highly relevant for the future developments of spin-based hybrid optomechanics~\cite{Ovartchaiyapong_NatureComm_14}. Along the same line, the strain response of dye molecules, which can also serve as local mechanical sensors~\cite{Tian_PRL_14}, could be reliably determined using this technique.

To conclude, we have shown that a hybrid core-shell nanowire geometry allows applying a uniform strain field on several QDs embedded in the core. Using this method, we have demonstrated that self-assembled InAs QDs elongated along their growth axis undergo large spectral shifts, with moderate dot to dot variations. Beyond their fundamental interest, our results provide a valuable input for the design of strained-QD devices, including emerging hybrid QD-optomechanical systems. Finally, this strategy could also be used to calibrate the strain response of other optically-active quantum emitters and/or individual spin hosted in such structures.

\section*{Methods}

\paragraph*{Sample fabrication.}

The sample is fabricated from a planar structure grown on a ($001$) GaAs wafer by Molecular Beam Epitaxy. Sample structure (from top to bottom): GaAs ($850\: $nm), self-assembled InAs QDs, GaAs ($850\: $nm), Al$_{0.85}$Ga$_{0.15}$As ($500\: $nm), GaAs buffer layer and wafer. The QDs are obtained by Stranski-Krastanov growth, through the deposition of $1.9$ monolayers of InAs at $530^{\circ}$C during $1\: $s (Ref.~\cite{Gerard_JCG_95}). After this step, the dots are immediately capped with GaAs.

Before top down definition of nanowires, the sample is flip-chipped and glued on a host substrate. This step was intended for other studies, and has no impact on the results presented in this work. The top part of the epitaxial sample is glued on a GaAs host substrate with an epoxy polymer. The growth wafer and the sacrificial Al$_{0.85}$Ga$_{0.15}$As layer are successively removed by mechanical and selective wet etching, leaving a mirror-flat surface. A nickel hard mask is then defined by electron-beam lithography, directive metal deposition and lift-off. The nanowires are etched in a Reactive Ion Etching chamber, using a SiCl$_4$-Ar gas mixture. The residual Ni mask is finally removed with diluted nitric acid. The sample consists in arrays of nanowires with nominal top diameters ranging from $200\: $nm to $800\: $nm.

The silica shell is deposited by Plasma Enhanced Chemical Vapor Deposition (PECVD) at a temperature of $280^{\circ}$C, using SiH$_4$ and N$_2$O gaz precursors and a RF power of $280\: $W. The deposition is roughly conformal: for a given sidewall thickness $t_s$, the vertical thickness is $1.5 \times t_s$.

\paragraph*{Numerical simulations.}

Mechanical simulations are conducted with the COMSOL software. The zincblende GaAs core is treated as a mechanically anisotropic material with the following  stiffness coefficients: $c_{11} = 122.1\: $GPa, $c_{12}= 56.6\: $GPa and $c_{44}= 60\: $GPa. The amorphous silica shell is treated as an isotropic material, characterized by its Young modulus $Y_{\text{SiO}_2} = 70\: $GPa and by its Poisson ratio $\sigma_{\text{SiO}_2} = 0.17$. The temperature dependence of these coefficients is neglected (we use $300\: $K values). The shell is supposed to be perfectly bonded to the core. Starting from the effective silica deposition temperature $T_d^\text{eff}$, the sample is cooled down to liquid helium temperature ($4\: $K). We take into account the temperature dependence of the linear thermal expansion coefficients of the two materials.

\paragraph*{Optical spectroscopy.}

Optical characterization is performed in a standard $\mu$PL setup. The sample is thermally anchored to the cold finger of a liquid helium cryostat (nominal temperature: $4\: $K) with an optical access. Optical excitation is provided by a Ti:sapphire laser focused on the sample with a microscope objective (numerical aperture: $0.6$). $\mu$PL QD spectra are obtained under pulsed excitation (repetition rate: $76\: $MHz) with a laser photon energy tuned to $1.534\: $eV, in the absorption continuum of GaAs bandgap. The same objective collects the QD luminescence signal; after filtering of residual laser stray light, the luminescence signal is sent to a grating spectrometer equipped with a silicon APD for spectral analysis.

A PL excitation technique is employed to measure the absorption spectrum of GaAs. In that case, the laser is operated in continuous mode and the photon energy is scanned while monitoring the intensity of a selected QD emission line. We have checked that the absorption features do not depend on the QD.

\section*{Acknowledgements}

The authors warmly thank Jo\"{e}l Cibert, Jean-Philippe Poizat, Armando Rastelli and Maxime Richard for fruitful discussions. We thank Marion Ducruet for the growth of the sample. Clean room processing was carried out in the `Plateforme Technologique Amont (PTA)' and CEA LETI MINATEC/DOPT clean rooms.

\section*{Funding}

The authors acknowledge the support of the European Union Seventh Framework Program 209 (FP7/2007-2013) under grant HANAS (601126 210), and of the French ANR with the grants XDISPE (ANR-11-JS10-004-01) and WIFO (ANR-11-BS10-0011).


\bibliography{core_v2}

\pagebreak

\begin{tocentry}

\includegraphics{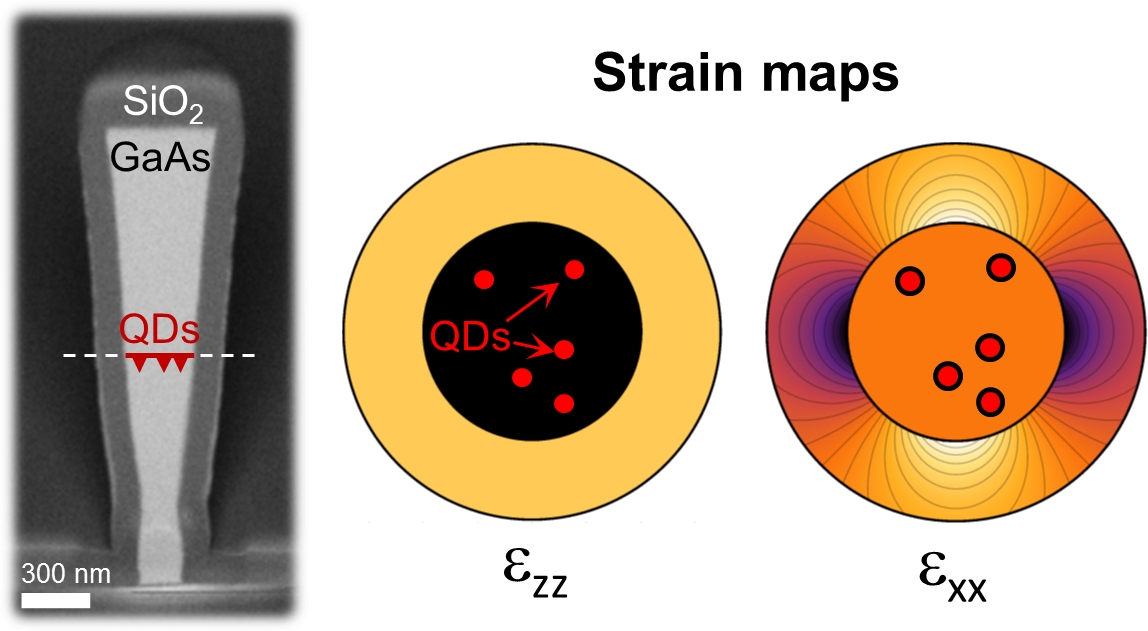}

\end{tocentry}

\end{document}